\documentclass[onecolumn,nofootinbib,superscriptaddress]{revtex4}

\usepackage{slashed}

\usepackage{slashed}

\usepackage[normalem]{ulem}

\def\({\left(}
\def\){\right)}
\def\[{\left[}
\def\]{\right]}
\usepackage{graphicx}

\usepackage{color}

\usepackage{epsfig}
\usepackage{epstopdf}
\usepackage{epsf}
\input{epsf.sty}
\usepackage{graphicx,amsmath,amssymb}
\usepackage{epsfig}
\allowdisplaybreaks

\def\ei{\end{itemize}}
\def\be{\begin{equation}}
\def\ee{\end{equation}}
\newcommand{\bea}{\begin{eqnarray}}
\newcommand{\eea}{\end{eqnarray}}

\usepackage{bbm,bm,amsmath,amssymb}

\usepackage{hyperref}

\newcommand{\kk}{\mathbf{k}}
\newcommand{\xx}{\mathbf{x}}
\newcommand{\yy}{\mathbf{y}}
\newcommand{\dd}[1]{\mathrm{d}#1\,}

\begin{document}

\setlength{\parskip}{2pt}
\title{V-mode Polarization in Axion Inflation and Preheating} 

\author{Stephon Alexander}
\email{stephon\_alexander@brown.edu}
\affiliation{Department of Physics, Brown University, Providence, RI, 02906}
\author{Evan McDonough}
\email{evanmc@physics.mcgill.ca}
\affiliation{Ernest Rutherford Physics Building, McGill University,\\ 
3600 University Street, Montr{\'e}al QC, Canada H3A 2T8\\}
\author{Robert Sims}
\email{robert\_sims@brown.edu}
\affiliation{Department of Physics, Brown University, Providence, RI, 02906}

\begin{abstract}
\vspace*{.5em}We study the production of primordial circular (``V-mode'') polarization in axion inflation coupled to fermions and gauge fields, with special attention paid to (p)reheating. We construct the power spectrum of $V$, and find a blue-tilted spectrum with index $n_V=4$. This is independent of the dominant decay channel of the inflaton (direct fermion vs. direct photon production).

\end{abstract}

\maketitle

\vspace*{-3em}

\tableofcontents

\vspace{8mm}

\section{Introduction}

One of the cornerstones of cosmic microwave background (CMB) physics is the prediction of linear polarization of the E (gradient) mode from the intrinsic quadrupole temperature anisotropy, induced by Thomson scattering, which is now measured in the CMB \cite{CMBpol}.  Inflationary models also predict B-mode (curl type) polarization which is generated from tensor metric perturbations and induces non-vanishing off-diagonal components of the polarization matrix.  This greatly increases the amount of `fundamental physics' that can be extracted from the CMB, for example, in the simplest models of inflation, a detection of primordial B-modes will probe the energy scale of inflation.

Another interesting piece of lore, which has not gotten as much attention, is the possibility that the CMB could have circular polarization present, otherwise known as a V-mode Stokes parameter,  which is expressed in terms of the photon polarization states as \cite{Finelli:2008jv} 
\begin{equation}
V = \frac{1}{a^4} \left( |A_+ ' |^2 - |A_- ' |^2\right) ,
\end{equation}
where $A_{\pm}$ are the two chiral polarizations of the photon, and $'$ is the derivative with respect to conformal time. The corresponding brightness temperature perturbation can then be constructed in an analogous manner to E and B \cite{Kosowsky:1994cy,Giovannini:2009ru}. The V-mode polarization is usually assumed to be zero because Thomson scattering (in the absence of a magnetic field) does not intrinsically source V. In the presence of a magnetic field, it was shown by Giovannini \cite{Giovannini:2009ru} that a non-vanishing 
$V$ can be produced.  One may wonder what else circular polarization may tell us about the early universe, especially the epoch of inflation.  Some aspects of this issue have been explored previously in e.g. \cite{Cooray:2002nm, Alexander:2008fp, Alexander:2016moy, Bavarsad:2009hm, Finelli:2008jv,   Giovannini:2009ru}.

The inflationary generation of circular polarization could be quite a general phenomenon provided that the inflaton field sources chiral symmetry breaking either directly in the photon sector or indirectly through coupling to fermions, and we will study both these scenarios. In the former case, there is a direct production of one polarization state during inflation, while in the latter case, the pseudoscalar sources a left-right asymmetry in a charged fermion which is subsequently transferred to circularly polarized photons \cite{Alexander:2016moy}. These mechanisms for generating CMB circular polarization are qualitatively different from the generation of E- and B-mode polarization, as well as the generation of $V$ by background magnetic fields \cite{Giovannini:2009ru}, as in the former case the polarization is generated \emph{during} inflation and reheating, while in the latter cases the polarization is only generated upon horizon re-entry of primordial scalar and tensor modes.

We will work in the context of axion inflation, supplemented by the couplings $\partial_\mu \phi J^{\mu5}$ and $\phi F \tilde{F}$, where $\phi$ is the axion, $J^{\mu5}$ is an axial fermion current, and $F$ is a gauge field strength tensor.  Models of axion inflation are particularly appealing due to the underlying shift symmetry of the axion, which prevents the $\eta$-problem of large field inflation, and due to their ubiquity in string theory (in addition to the historical significance of the axion as a solution to the strong CP-problem of the standard model). We will we compute the power spectrum of V-mode anisotropies at the end of inflation and (p)reheating, leaving the full evolution to last scattering for future work.  Remarkably, we find that both interactions lead to the same spectrum of V-mode anisotropies, which is blue-tilted with a spectral index $n_V=4$.

For both the couplings $\partial_\mu \phi J^{\mu5}$ and $\phi F \tilde{F}$, the physical origin of non-zero V is the definite sign of $\dot{\phi}$ during inflation, which produces a net circular polarization on super-Hubble scales.  However, after inflation, the field $\phi$ oscillates and both polarizations are produced, making the predictions for V-mode polarization sensitive to the detailed dynamics of preheating.  This is in contrast to most other CMB observables, for example $n_s$ and $r$, which are largely decoupled from the details of (p)reheating. The preheating of gauge fields via $\phi F \tilde{F}$ has been studied in detail in \cite{Adshead:2015pva} and \cite{McDonough:2016xvu} (see also \cite{Adshead:2016iae}), and in fact the requisite ingredients to construct $V$ are already contained in \cite{McDonough:2016xvu}.  Meanwhile the preheating production of neutral fermions via $\partial_{\mu}\phi J^{\mu 5}$ has been studied in \cite{Adshead:2015kza}, which we extend to study charged fermions\footnote{Note that our analysis differs from that of \cite{Adshead:2015jza}, which studied Majorana fermions. These are not useful for generating gauge fields, since the vector current $J^\mu$ vanishes identically for Majorana fermions.}. In a more realistic setup, these two couplings will both be present and will be competing channels for preheating. For simplicity, we will consider them separately.  We comment on the competition between these two couplings in Appendix B.

The outline for this paper is as follows: in Section II we discuss preliminaries of circular polarization and introduce the quantities we will compute. In Section III we present the relevant equations of motion and describe the dynamics during inflation of this model, followed by an analysis of preheating in Section IV, and in section V we compute the VV power spectrum on large scales. We close in Section VI with a discussion of prospects for detection and directions for future work.

\section{Circular Polarization Preliminaries}

We have expressed $V$ in equation (1) as a difference in the photon polarization states; this will be the most useful definition for our analysis. This definition is related to the more conventional definition $V = 2 \, \mbox{Im} [ E_x ^{*} E_{y}] $ by the change of basis to $\{x_+,x_-\}$ coordinates, $\sqrt{2} \hat{x}_+ = \hat{x} + i \hat{y}$, $\sqrt{2} \hat{x}_- = \hat{x} - i \hat{y}$ \footnote{In this basis, $\sqrt{2} E_x = E_+ + E_-$ and $\sqrt{2} E_y = i(E_+ - E_-)$, which gives $V = 2 \, \mbox{Im} [ E_x ^{*} E_{y}] = E_+ ^2 - E_- ^2 $. In an FRW spacetime, this is precisely equation (1).}. This can also be expressed in terms of components of the `polarization matrix' as
\be
V = - i( \rho_{12} - \rho_{21})
\ee
where $\rho_{ij}$ is the polarization matrix defined by \cite{Kosowsky:1994cy}
\bea
\rho =  
&&  \frac{1}{2}\begin{pmatrix}
  I + Q  & U - i V  \\
  U + i V & I - Q  \\
  \end{pmatrix} \\
  = && I \, \mathbb{I} + Q \, \sigma_3 + U \, \sigma_1 + V \sigma_2 \, ,
\eea
where in the second line we have used the Pauli matrices $\sigma_i.$

The $V$ defined above has units of intensity. Anisotropies in $V$ can be converted to a fractional temperature fluctuation, which we denote $\Theta_V$, via the rescaling \cite{Lyth:2009zz,Hu:2002aa,calPaper},
\be
\label{deltaTV}
\Theta_V \equiv \frac{\delta V_T}{T} = \frac{\delta V}{I} ,
\ee
where $V_T$ is the V stokes parameter in units of temperature, $I\equiv\frac{1}{a^4} \left( |A_+ ' |^2 + |A_- ' |^2\right)$ is the intensity stokes parameter, and $T$ is the background blackbody temperature. One can then construct the $C_l ^{VV}$ as the coefficients in the multipole expansion of the two-point function $\langle  \delta V_T \delta V_T\rangle$.

The quantity we compute in this paper is the polarization present at the end of inflation and reheating, which serves as the initial condition for the subsequent evolution to last scattering. In the case of $E$ and $B$, the initial polarization is ignored, for good reason: it is heavily suppressed by scatterings and is negligible compared to the polarization produced by primordial scalar and tensor perturbations. In contrast, for $V$ (in the absence of a magnetic field) there is no signal generated by primordial scalar and tensor fluctuations, so the only inflationary $V$ will be a relic of that produced \emph{during} inflation. However, a mechanism is still required to circumvent the suppression from scatterings. We will not fully develop this mechanism here, but will discuss possibilities in Section \ref{sec:Discussion}.

Analogous to the primordial scalar spectrum, the (dimensionless) primordial power spectrum of V-mode polarization is given by
\be
P_{\Theta_V} (k) = \frac{1}{I^2} \frac{k^3}{2 \pi^2} |\delta V_k|^2 . \label{PowSpec}
\ee
This can be parametrized in a similar way to the primordial scalar spectrum, see e.g. \cite{Ade:2015lrj}, as
\be
\label{PThetaVForm}
P_{\Theta_V} (k) = \mathcal{A}_V \left(\frac{k}{k_0}\right)^{n_V - 1}
\ee
where $\mathcal{A}_V$ is the amplitude of primordial V-mode anisotropies at a reference scale $k_0$, and $n_V$ is the spectral index. 


\section{Background Dynamics: axion inflation}

We will study axion inflation coupled to a low energy U(1) gauge-fermion system with a triangle anomaly cancelling term. The action for this model is given by\footnote{Using metric signature $(-,+,+,+)$.}   
\begin{eqnarray}
S = \int {\rm d}^4 x \sqrt{- g} \left[ \frac{M_{\rm Pl}^2}{2}R - \frac{1}{2}(\partial\phi)^2 -  V(\phi) + \overline{\psi}(i \gamma^\mu D_{\mu} - m_\psi) \psi -\frac{1}{4}F_{\mu\nu}F^{\mu\nu} + A_{\mu} J^{\mu} + \frac{C}{f} \partial_{\mu}\phi J^{\mu 5}  +
\frac{\alpha}{f} \phi F_{\mu\nu}\tilde{F}^{\mu\nu} \right].
\end{eqnarray}
In the above, $F_{\mu \nu}$ is the usual field strength tensor of the photon, $D_\mu$ is the covariant derivative with respect to the spin connection, and the vector current $J_{\mu}$ and axial vector current $J^{\mu 5}$ are given by
\begin{equation}
J^{\mu} = g \, \overline{\psi} \gamma^{\mu} \psi \;\;\; , \;\;\; J^{\mu 5} = \overline{\psi} \gamma^{\mu} \gamma_5 \psi . 
\end{equation}
The fermion $\psi$ is a 4-component Dirac spinor charged under the standard model gauge group, though we will only consider the effective coupling to $U(1)_{\rm EM}$.

The background cosmology of this model is dictated by the Friedman equation,
\begin{equation}
H^2 =  \frac{1}{3 M_{Pl}^2} \left[ \rho_{\phi} + \rho_{A} + \rho_{\psi} \right] ,
\end{equation}
where $\rho_\phi$ is the inflaton energy density, which we will assume is dominant during inflation, while $\rho_{A}$ and $\rho_\psi$ are the effective background energy density in the gauge field $A_\mu$ and the fermion $\psi$ respectively. This is in addition to the equation of motion of the inflaton, given by
\be
\label{phiEOM}
\ddot{\phi} + 3 H \dot{\phi} + V_{,\phi} = - \frac{C}{f} \partial_\mu J^{\mu 5} + \frac{\alpha}{f}  F_{\mu\nu}\tilde{F}^{\mu\nu} . \,
\ee
While the mechanism we consider here is independent of the choice of inflationary potential, for concreteness we will consider the classic example of
\be
V(\phi) = \frac{1}{2} m_\phi ^2 \phi^2 ,
\ee
with a benchmark value for the mass of $m_\phi = 10^{-6} M_{Pl}$. Such a potential for an axion arises in monodromy models, e.g. the F-term axion monodromy model of \cite{Marchesano:2014mla}. 

Provided that backreaction is not significant during inflation, such that we can ignore the source terms on the RHS of \eqref{phiEOM}, inflation ends once the slow-roll conditions are violated. For $m^2 \phi^2$ inflation this occurs at (see e.g. \cite{Baumann:2009ds})
\be
\phi_{end} = \sqrt{2}M_{Pl}.
\ee
At this point $(1/2) \dot{\phi}^2 = M_{Pl} ^2 H^2$, which follows from $\epsilon = (1/2) \dot{\phi}^2 / (M_{Pl} ^2 H^2)$, and hence the value of $\dot{\phi}$ is given by
\be
\dot{\phi}_{end} = m_\phi M_{Pl}=\frac{1}{\sqrt{2}}m_\phi\phi_\text{end}.
\ee
After inflation, and in the absence of expansion and backreaction, the field $\phi$ begins to oscillate, triggering the `preheating' phase. In this phase the field is described by, 
\be
\phi(t) = \phi_{end} \sin(m_\phi t) .
\ee
The maximum field velocity in this phase is thus $\dot{\phi}_{pre}= m_\phi \phi_{end}$, which is roughly a factor of $\sqrt{2}$ larger than the maximum value of the field velocity during inflation.

During inflation there will be production of fermions and gauge fields due to the interactions in the Lagrangian. The relative strength of these interactions is controlled by the ratio of parameters $C/\alpha$. We will consider the regimes $\left|C/\alpha\right| \gg1$ and $\left|C/\alpha\right| \ll 1$ separately.

\subsection{Charged Fermion and Gauge Field Production during Inflation with $\partial_\mu \phi J^{\mu5}$}

In the region of parameter space $\left|C\right| \gg \left|\alpha\right|$, the dominant interaction for $\phi$ is with fermions. To describe the fermion dynamics, it will be convenient to decompose the 4-component Dirac spinor $\psi$ into two 2-component Weyl spinors $\varphi$ and $\eta$, via
\begin{equation}
\psi = {\varphi \choose \eta^\dagger} ,
\end{equation}
in terms of which the fermion currents take the form
\begin{equation}
J^{\mu} = g \left( \varphi^\dagger \overline{\sigma}^\mu \varphi -  \eta^\dagger \overline{\sigma}^\mu \eta \right) \;\;\; , \;\;\; J^{\mu 5} = \varphi^\dagger \overline{\sigma}^\mu \varphi +  \eta^\dagger \overline{\sigma}^\mu \eta . 
\end{equation}
Working in the comoving time FRW metric, we can reduce the covariant derivative to a partial derivative by rescaling the fermion fields by $a^{-3/2}$ to absorb the factor of $\sqrt{-g}$, as in \cite{Adshead:2015kza}.  The fermionic action then takes the form:
\begin{align} 
S_{\rm f} =  \int {\rm d}^4 x \left[ i \varphi^{\dagger} \bar{\sigma}^{\mu }\partial_{\mu} \varphi +i \eta^{\dagger} \bar{\sigma}^{\mu }\partial_{\mu} \eta  \right. & - m_\psi(\varphi  \eta + \varphi^{\dagger} \eta^\dagger )   + \frac{C}{f}\partial_{\mu}\phi( \varphi^{\dagger}  \bar{\sigma}^{\mu} \varphi + \eta^{\dagger} \bar{\sigma}^{\mu } \eta) + g A_{\mu} (   \varphi^\dagger \overline{\sigma}^\mu \varphi -  \eta^\dagger \overline{\sigma}^\mu \eta ) \left. \right] .
\end{align}

Neglecting the $\phi F \tilde{F}$ interaction, the dynamics of the gauge-fermion system are dictated by the gauge field equation of motion,
\be
\partial_{\nu} \left( \sqrt{-g} F^{\mu \nu}\right) = J^{\mu} ,
\ee
and the fermion equation of motion, 
\begin{align}
i\bar{\sigma}^{\mu }\partial_{\mu} \varphi + \left( \frac{C}{f}\partial_{\mu} \phi \,   + g  A_{\mu}\right) \bar{\sigma}^{\mu} \varphi  & = m_\psi \eta^{\dagger } ,\\ 
i\bar{\sigma}^{\mu }\partial_{\mu} \eta + \left( \frac{C}{f}\partial_{\mu}\phi - g  A_{\mu} \right) \bar{\sigma}^{\mu } \eta  & = m_\psi\varphi^\dagger  .
\end{align}
During inflation the time-variation of $\phi$ leads to a violation of adiabaticity for the fermions, leading to non-perturbative particle production wherein one helicity of the fermions is preferentially produced \cite{Adshead:2015kza}. The results of \cite{Adshead:2015kza}, which did not include gauge fields, apply to our case at times when gauge field production on the fermion equation of motion is negligible, or more precisely $\frac{C}{f} \dot{\phi} \gg g A_0 \, , \, g A_i .$ This assumption eventually breaks down and the analysis must be done numerically. For our purposes, we will use their results for the fermion production and study the corresponding gauge field production.

The fermions can be expanded in mode functions as
\begin{eqnarray}
\varphi_\alpha(x, t) = && \sum_{\lambda} \int \frac{{\rm d}^3 k}{(2\pi)^3}\left[x^\lambda_{\alpha k}( t) a_{ k}^\lambda e^{i { k}x} + y^\lambda_{\alpha k}( t) b^{\lambda\dagger}_{ k} e^{-i { k}x}\right], \nonumber \\
\eta_\alpha({\bf x}, t) = && \sum_{\lambda} \int \frac{{\rm d}^3 k}{(2\pi)^3} \left[x^\lambda_{\alpha k}( t) b_{k}^\lambda e^{i k x} + y^\lambda_{\alpha k}( t) a^{\lambda\dagger}_{ k} e^{-i {k}x}\right],
\end{eqnarray}
and further decomposed into a helicity basis via the definition
\be
x^{\lambda}_k( t) =  X^{\lambda}_{k} (t) \xi_\lambda({\bf k}) \;\;,\;\; y^{\lambda \dagger } _k ( t) =  Y^{\lambda *}_{ k}(t) \xi_\lambda({\bf k}).
\ee
where $\lambda=\pm$ denotes the (+) and (--) helicity states. Explicit expressions for the helicity eigenspinors $\xi_\lambda$ can be found in \cite{Koksma:2009tc}. 

The particle number is then defined in terms of these mode functions as \cite{Adshead:2015kza},
\be
n_k ^{\lambda} = \frac{1}{\omega_{\lambda} (\tilde{k}_{\lambda} + \omega_{\lambda})} \left[ |\dot{X}_k ^\lambda|^2 + \omega_{\lambda}^2 |X_k ^{\lambda}|^2 - 2 \omega_{\lambda} \; \mbox{Im}( X_k ^{\lambda} \dot{X}_k ^{\lambda}) \right] ,
\ee
where the modified dispersion relations are,
\be
\label{disprel}
\omega_{\lambda}^{2}(t) = \tilde{k}_{\lambda}(t)^{2} + m_\psi ^{2},	 \; \;  \tilde{k}_{\lambda}(t) = \left(\frac{k}{a}\lambda + \frac{C}{f} \dot{\phi} \right) .
\ee
After matching to the Bunch-Davies vacuum, the mode functions have the form\footnote{Where we have corrected for a typo in the normalization stated in \cite{Adshead:2015kza}.},
\begin{eqnarray}\label{eqn:inflationsols}
X^+_{ k}(k\tau) =    - \frac{i m_\psi}{H}  \frac{e^{i\theta}e^{\frac{\pi}{2}\vartheta}}{\sqrt{2k\tau}}W_{-\frac{1}{2}-i\vartheta,\mu}(2 i k\tau),&& \quad X^-_{ k}(k\tau) =   \frac{e^{i\theta}e^{-\frac{\pi}{2}\vartheta}}{\sqrt{2k\tau}}W_{\frac{1}{2} + i \vartheta, \mu}(2 i  k\tau) ,\nonumber \\
Y^{+*}_{k}(k\tau) = \frac{e^{i\theta'}e^{\frac{\pi}{2}\vartheta}}{\sqrt{2k\tau}} W_{\frac{1}{2}-i\vartheta, \mu}(2 i  k\tau),\qquad\quad&& \quad Y^{-*}_{ik}(k\tau) = -\frac{i m_\psi}{H}  \frac{e^{i\theta'}e^{-\frac{\pi}{2}\vartheta}}{\sqrt{k\tau}} W_{-\frac{1}{2}+ i \vartheta, \mu}(2 i  k\tau),
\end{eqnarray}
where $W_{x,y}(z)$ are Whittaker functions, and we have defined,
\be
\vartheta = - \frac{C}{f} \frac{\dot\phi}{H} \;\;,\;\; \mu^2 =  -\( \frac{m_\psi ^2}{ H^2} + \vartheta^2\),
\ee
and $\theta$, $\theta'$, are arbitrary phases. The particle number on large scales is then given by
\be
n_{k}^{\pm} = \,e^{-\pi \left( \mp\vartheta + \sqrt{\frac{m_\psi ^2}{H^2}+\vartheta^2}\right) } \frac{\sinh\left[\pi \left( \sqrt{\frac{m_\psi ^2}{H^2}+\vartheta^2}\pm\vartheta \right)\right]}{\sinh\left[2\pi\left( \sqrt{\frac{m_\psi ^2}{H^2}+\vartheta^2}\right)\right]}.
\ee
At strong coupling $\vartheta \gg m_\psi/H$ this simplifies to
\be
n_k ^+ \approx 1 \;\;,\;\; n_k ^-  \approx 0 .
\ee
There is thus a large asymmetry in the helicity states. This is similar to inflation with the coupling $\phi F \tilde{F}$, wherein one polarization of the gauge fields is amplified and other is negligible.

Finally, the helicity asymmetry in the fermions will be transferred to the photon via perturbative processes, namely Bremsstrahlung. In the case of single-Bremsstrahlung, this process allows a (+1/2) spin fermion to convert to a (--1/2) fermion via the emission of near-collinear (+1) spin photon, and a (--1/2) spin fermion will convert to a (+1/2) fermion via the emission of near-colinear (-1) spin photon. In our case, inflation and preheating will produce a large number of (+) helicity fermions, leading to production of $(+)$ photons. The modern theoretical framework to describe this process is the spinor-helicity formalism for gauge theories, as reviewed in e.g. \cite{Dixon:2013uaa} \cite{Schwartz:2013pla}. The emission of Bremsstrahlung in this framework was first studied in \cite{DeCausmaecker:1981jtq, Berends:1981uq}, where the amplitudes for all relevant processes were computed. In our work we will take $m_{\psi}/H$ small but finite, such that helicity is approximately conserved on large scales and $V/I \sim \mathcal{O}(1)$ for the produced photons\footnote{Where $I$ is the intensity stokes parameter, which determines $T_{\mbox{CMB}}$ and the TT power spectrum.}.

\subsection{Gauge Field Production during Inflation from $\phi F \tilde{F}$}

If instead $\left|\alpha\right| \gg \left|C\right|$, the dominant interaction term is that between the inflaton and the gauge field, $\phi F \tilde{F}$. This mechanism for the production of gauge fields from the coupling  has been considered in many works, e.g. \cite{Anber,Barnaby, Adshead:2015pva}. The equation of motion for $A_{\mu}$ is
\be \label{EoMA}
\frac{d^2 {A}_{k\pm}}{d \tau^2} + \left( k^2 \pm 2k \frac{\xi}{\tau} \right) A_{k\pm} \, = \, 0 \, , 
\ee
where $\xi$ is given by
\begin{equation} 
\xi \, = \, \frac{2 \alpha \dot{\phi}}{f H} \, .
\end{equation}
The parameter $\xi$ plays a similar role to $\vartheta$ in the fermionic case, and there is production of one of the polarization states on scales with $k$ less than a critical value set by $\xi$. In this case, $A_{k+}$ modes which satisfy
\be
\frac{k}{aH} < 2 \xi ,
\ee
experience a tachyonic instability and are amplified during inflation, while $A_{k-}$ is unaffected. The mode function prepared by inflation is
\bea
\label{modefunctioninflation}
{A_{k+} ^{(0)}}  \, &=& \, 
\frac{2^{-1/4}}{\sqrt{2k}} \left( \frac{k}{ \xi a H}\right)^{1/4} e^{\pi \xi - 4 \xi \sqrt{ k / 2 \xi a H}} \nonumber \\
{A_{k-} ^{(0)}} \, &\approx& \, 0 ,
\eea
where the $+$ mode on large scales is amplified by a factor of $e^{\pi\xi}$.

Current CMB observations bound the value of $\xi$ at the moment the CMB pivot scale $k_{\star}$ exits the horizon to be $\xi_{\star} \leq 2.2 $ \cite{Ade:2015lrj}, which corresponds to a (model-dependent) bound on the coupling $(\alpha/f)\lesssim 110 M_{Pl}^{-1} - 125 M_{Pl} ^{-1}$ for $m^2 \phi^2$ inflation \cite{Adshead:2015pva}. When we refer to the `strong coupling' regime of this model, we mean the range $1 M_{Pl} ^{-1} \lesssim (\alpha/f) \lesssim \mathcal{O}(10^2) M_{Pl} ^{-1}$.

\section{Preheating}

After inflation, the oscillatory behaviour of the inflaton can lead to instabilities and explosive particle production for fields directly coupled to the inflaton. This phenomenon is known as ``preheating", originally discovered in \cite{STB,STB2,KLS,KLS2}, and more recently reviewed in  \cite{ABCM,Karouby}. As hinted at in the introduction, the predictions for $V$ can be very sensitive to the details of preheating. With this in mind, we will undertake an analysis of preheating which seeks to uncover the extent to which circular polarization produced during inflation will survive preheating.

For the case of the direct coupling $\phi F \tilde{F}$ between the axion and gauge fields, the preheating dynamics are straightforward. It was shown in \cite{Adshead:2015pva, McDonough:2016xvu} that gauge fields are copiously produced and preheating terminates quickly provided that the coupling is sufficiently large. Preheating into fermions via Yukawa couplings was originally studied in \cite{Greene:2000ew}, and subsequently analyzed in many works e.g. \cite{Adshead:2015kza,FermionicPreheating}. However, preheating into charged fermions via $\partial_\mu \phi J^{\mu 5}$, which then produces photons, is more subtle, and has not been studied thus far. We dedicate the following section to this issue. That is, we will work in the regime,
\be
\left|\frac{C }{\alpha}\right| \gg 1 .
\ee

\subsection{The Structure of Preheating into Charged Fermions}

In this section, the basic mechanism we would like to consider is the non-perturbative production of fermions, which is instantaneous and occurs once an inflaton oscillation, and the subsequent perturbative production of photons. The simplest scenario is that preheating terminates after one half-oscillation of the inflation, such that $\dot{\phi}$ never switches sign, and the maximal helicity asymmetry of fermions, and consequently circularly polarization of photons, is achieved. 

We will show that this occurs provided that the requisite `new physics' (as measured by $(C/f)^{-1}$) occurs near the GUT scale, but at a sufficiently high scale that backreaction does not prevent preheating from occurring.  For smaller values of the coupling (i.e. a higher energy scale for new physics), preheating lasts for multiple or many cycles allowing for production of both helicity states, which suppresses the circular polarization.  In all cases, the conversion to photons then takes place via perturbative processes, occurring within a single Hubble time. Perturbative reheating continues after this point, operating on sub-Hubble scales, until the universe reaches near-thermal equilibrium and the radiation phase of standard big bang cosmology begins.

Before we proceed with preheating, let us recall that if the dominant interaction is between the gauge field and fermion current, the general solution of the gauge field is:
\begin{equation}
A^{\pm}_k(\tau) \sim i\int\frac{\dd{\eta}}{a(\eta)}G_k(\eta,\tau)J^{\pm}(X^{\lambda}_{k}, Y^{\lambda}_{k}) ,
\end{equation}
where $J^{\pm}$ is the $(+/-)$ helicity piece of the vector current, and $G_k$ is the Green's function of $A_k$.
We can define the relative chirality to be 
\be
\frac{A_{+} - A_{-}}{A_{tot}} = A_{rel}.
\ee
So as long as there is some linear polarization present the total amount of gauge fields will be non-vanishing, $A_{tot} \neq 0$.  We can express the relative photon chirality as:
\be
A_{rel}(X^{\lambda}_{k}, Y^{\lambda}_{k};\tau) =  i A_{tot}^{-1}\int\frac{\dd{\eta}}{a(\eta)}G_k(\eta,\tau)[J_{k}^{+}- J_{k}^{-}].
\ee
Note that during preheating, we can construct the chiral currents $J^{\pm}$ as a quadratic form of the eigenmodes produced during preheating, $ X^{\lambda}_{k}$ and $Y^{\lambda}_{k}$.  
We immediately see that if inflation produced a preponderance of left handed photons, then as long as the difference between left and right handed current are $\mathcal{O}(1)$ of the total current, then the chirality of the photons will be non-vanishing. Therefore under reasonable assumptions, the backreaction of the fermion production during preheating will not wash out the initially large photon helicity produced during inflation from potential lepton chirality flipping transitions.  To get an explicit computation of the percentage of chirality that is retained during reheating detailed numerical analysis is necessary and we plan to pursue this in a future work.

The structure of preheating is revealed by comparing time-scales in the problem. The time-scale for fermion production is the oscillation period of the inflaton, which is smaller than the Hubble time by roughly a factor of 10. Meanwhile, the time-scale for the production of photons is given by,
\be
\tau_{\gamma} = 1/\Gamma_{\gamma}
\ee
where $\Gamma_{\gamma}$ is the usual rate of QED-like interactions at finite temperature, given by
\be
\Gamma = n \sigma v  = g^2 T, 
\ee
where $T$ is the effective temperature of the QED-like sector, which is roughly given by $T \approx \rho_{\psi} ^{1/4}$. The time-scale for production of photons is then given by
\be
\tau_{\gamma} = \frac{1}{g^2 \rho_{\psi}^{1/4}} .
\ee
The relevant scale for comparison is the Hubble-time, which after the first production event is given by $\tau_{H} = M_{Pl}/\sqrt{\rho_{\psi}}$. Hence the ratio of the time-scales is given by
\be
\frac{\tau_{\gamma}}{\tau_{H}} = \frac{\rho_\psi ^{1/4}}{g^2 M_{Pl}} = \mathcal{O}(1) ,
\ee
where the second equality follows from the expressions and numerical values already used, in addition to $g \sim \alpha_{EM}\sim 10^{-2}$.  Thus the production of photons takes place in roughly a Hubble time after the fermions are produced and preheating is terminated. Smaller values of $g$ will lead to a longer time-scale for photon production, which will not substantially alter the structure of preheating.

With this in mind, we now study the production of fermions during preheating. The inflationary solution for the fermion mode functions is no longer valid during preheating, as the background is no longer adiabatically varying. These solutions were studied in the past and we will make some general remarks about the following WKB solutions for the different helicity eigenmodes:
\be
X^{\lambda}_{k}(t)=  \sqrt{1+\frac{\tilde{k}_{\lambda}}{\omega_{\lambda}}}e^{i\int \omega_{\lambda} dt} \;\; , \;\; Y^{\lambda}_{k}(t) = -\sqrt{1- \frac{\tilde{k}_{\lambda}}{\omega_{\lambda}}}e^{i\int \omega_{\lambda} dt}
\ee
where $\tilde{k}_{\lambda}$ and $\omega_{\lambda}$  are given in equation \eqref{disprel}.

During preheating the inflaton field oscillates about its potential minimum and adiabaticity can be violated. A simple calculation reveals that this occurs when the effective wave-number $\tilde{k}_\lambda$ vanishes
\be
\frac{k}{a}\lambda + \frac{C}{f} \dot{\phi}=0 .
\ee
Adiabaticity is violated for every $k$-mode twice an oscillation, once when $\dot{\phi}$ is positive (which produces (+) helicity fermions) and once when $\dot{\phi}$ is negative (which produces (--) helicity fermions). This violation of adiabaticity leads to the production of particles. 

The fields produced by preheating depend sensitively on the time at which preheating ends. This occurs once the `preheat fields' disrupt the inflaton equation of motion or else become comparable in energy density to the inflaton and thus take over the background dynamics. This provides two conditions for non-termination of preheating,
\be
\label{cond1}
V_{, \phi} \gg - \frac{1}{a^3} \, \frac{C}{f} \langle \partial_\mu J^{\mu 5} \rangle ,
\ee
and
\be
\label{cond2}
\rho_{\phi} \gg \rho_{\psi} + \rho_{A} .
\ee
We will focus on the second condition, as this suffices to provide a lower bound on $C/f$ such that preheating terminates after one production event (i.e. one half-oscillation). 

The energy density in fermions is given by
\be
\rho_{\psi}  = \sum_{\lambda} \int \mathrm{d}^3 k \; n^\lambda _k \omega_\lambda (k) ,
\ee
where $n_k ^\lambda$ is the number density and $\omega_\lambda (k)$ is the energy-per-particle. After one production event, and before conversion into photons, the number density in (--) helicity states vanishes while the number density in (+) helicity states is given by
\begin{align}
\label{numberdensitypreheating}
n_k^{+} =  \left\{ \begin{array}{ll}\exp\left( -\pi{ m_\psi \over  \sqrt{k_c ^2-k^2} }\right ) & \,, \, k < k_c \\ 0 & , k> k_c. \end{array}\right.
\end{align} 
where the `critical wave number' $k_c$ is defined by
\be
 k_c \equiv \frac{C}{f} |\dot{\phi}_{pre}| .
\ee
The same number density applies at the end of inflation, with $\dot{\phi}_{pre}$ being replaced by $\dot{\phi}_{end}$.

The above expressions simplify in the limit $\vartheta \gg m_{\psi}/H$. In this case, the energy density in fermions can be computed explicitly and is given by
\be
\rho_{\psi} = \frac{\pi k_c^4}{3}\left(1-\pi (3\pi-8)\frac{m_\psi}{k_c}+\mathcal{O}(m_\psi/k_c)^{3/2}\right),
\ee 
where terms of $\mathcal{O}(m_\psi/k_c)$ will be neglected. Meanwhile the energy density in the inflaton during preheating is given by
\be
\rho_{\phi} \approx  \frac{3}{4}m_\phi^2\phi_\text{end}^2.
\ee

One can now easily compute the lower bound on $C/f$ such that backreaction does not shut off preheating before it begins. Using the value of $k_c$ during inflation, the condition \eqref{cond2} can be rewritten as a constraint on $C/f$, as
\be
\frac{C}{f} < \frac{1}{\sqrt{m_\phi \phi_{end}}}\left(\frac{9}{\pi}\right)^{1/4}.
\ee
For $C/f$ violating this bound, backreaction is already significant during inflation and preheating does not occur. During the first half-cycle of preheating the critical wave number changes by a factor of $\sqrt{2}$, which modifies this condition to
\be 
\frac{C}{f} < \frac{1}{\sqrt{m_\phi \phi_{end}}}\left(\frac{9}{4\pi}\right)^{1/4}.
\ee
For $C/f$ violating this bound, preheating terminates after one production event. Putting in the canonical values for $m$ and $\phi$, and re-interpreting $C/f$ as a scale of new UV physics, $C/f \equiv 1/\Lambda_{UV}$, we then find that preheating will terminate either before or immediately after one production event provided $\Lambda_{UV}$ is below an upper bound given by
\be
\label{StrongCoupling}
\Lambda_{UV} < 10^{-3}M_{Pl}\sim 10^{15}\text{ GeV}.
\ee 
In this regime there are no (--) helicity fermions produced, giving a maximal helicity asymmetry. This can be rephrased as the condition $\vartheta > 10^{3}$ during preheating.

In the opposite regime, $\Lambda_{UV} > 10^{15}$ GeV, preheating lasts for many cycles and the number density is modified from the expression \eqref{numberdensitypreheating}. The key difference from the previous regime is that there is now a production of (--) helicity fermions, and hence gauge fields, which occurs when $\dot{\phi}$ is negative. In this case, the expansion of the universe causes $k_c$ to redshift, which not only changes the maximum $k$ which is populated but also decreases the efficiency of particle production on large scales. 

A thorough study of this regime must rely on numerics, as was done by \cite{Adshead:2015kza}. However, we can make some analytic progress. The particle number on large scales following the $i^{th}$ production event is roughly
\be
n^{i} _k = e^{- \pi \frac{m_\psi}{k_c (t_0)} \left(\frac{a(t_i)}{a(t_0)}\right)^{3/2}} ,
\ee
where the helicity ($\pm$) of the produced particles is dictated by the sign of $\dot{\phi}$ at the $i^{th}$ event. The impact of the redshift factor, and the remaining helicity asymmetry on large scales $k\ll k_c$, depends sensitively on the ratio $m_\psi/k_c$. 

For $\Lambda_{UV} > 10^{15} {\rm GeV} $ and $m_\psi/k_c \ll 1$, the redshift factor $\left(a(t_i)/a(t_0)\right)^{3/2}$ (which for the first complete oscillation is roughly 2) is irrelevant and the production of (--) fermions is just as efficient as the production of (+) fermions. The residual asymmetry present after the subsequent oscillations is thus expected to be small, though a quantitative estimate requires numerical analysis. If instead $m_\psi/k_c \sim \mathcal{O}(1)$, the production of (--) is much less efficient but the initial $n^+ _k$ is only $e^{- \pi}\sim .04$. From this we conclude that the regime $\Lambda_{UV} > 10^{15} {\rm GeV} $ will not lead to a helicity asymmetry on large scales. 

There does however remain a spatially averaged \emph{net} helicity asymmetry in this regime, as computed in  \cite{Adshead:2015kza}, which occurs due to modes with $k \sim k_c$ at the beginning of preheating which decouple once $k_c$ becomes larger than $k$. However, for $\Lambda_{UV} \sim 10^{-4} M_{Pl}$ these modes are on a much smaller length scale than is of interest for CMB observations.

\section{The Spectrum of Circular Polarization on Large Scales}
\label{spectrum}
Now we come to the primary goal of this paper: to compute the large scale circular polarization, and in particular, the spectrum.  
For both production channels we work in the `strong coupling regime', such that preheating terminates before any (--) helicity particles can be produced.

\subsection{Indirect production via $\partial_\mu \phi J^{\mu 5}$}

This computation of $V$ is in principle a tedious calculation involving integrals over fermion mode functions (which we indeed compute in Appendix \ref{appVk}), but there is a intuitive shortcut that can be used to extract the spectral tilt of the V-mode spectrum: provided that the helicity asymmetry in the fermions is efficiently transferred to the gauge field, then the energy density in the gauge fields is precisely equal to the V-mode polarization, i.e.
\be
\rho_A = |\dot{A}_+|^2 + |\dot{A}_-|^2 \approx |\dot{A}_+|^2 ,
\ee 
and
\be
V = |\dot{A}_+|^2 -  |\dot{A}_-|^2 \approx |\dot{A}_+|^2 = \rho_A .
\ee
Moreover, at linear order in energy density perturbations and metric perturbations, and provided the energy transfer from fermions to photons is via perturbative processes (as opposed to, say, a parametric resonance instability), the spectrum of energy density fluctuations $\delta \rho$ will be unchanged as energy is transferred from the fermions to the gauge fields.  This follows from the lack of mode-mixing in linear perturbation theory. It follows from this that (up to an overall normalization) we can equate the Fourier modes of the energy density in fermions and gauge fields  i.e.
\be
\delta \rho_{\psi k} \propto \delta \rho_{A k} ,
\ee 
where the proportionality is up to a time-dependent normalization describing the transfer of energy from the fermions to gauge fields.

The spectrum of fermion energy density fluctuations is encoded in the number density and effective frequency, as the fermion energy density in a given Fourier mode is, up to a random phase, given by
\be
\delta \rho_{\psi k} = \sum_{\lambda} n_{k \lambda} \omega_{k\lambda} .
\ee 
As per our previous discussions, the number density at large coupling and on large scales is $k$-independent, as is $\omega_{k\lambda} \sim (C/f) \dot{\phi}$. From this it follows that $ |\delta \rho_{A k}|^2$ on large scales is independent of $k$, and the V-mode Fourier modes are given by
\be
|\delta V_k| = \mathcal{N} ,
\ee
for a time-dependent constant $\mathcal{N}$. This result is confirmed via explicit computation in Appendix \ref{appVk}, where we find the result,
\be
|\delta V_k|^2 =  \frac{16 g^4 f_h^2}{a^8(\tau)}  \left( \vartheta a H\right)^9  \mathcal{I}(\tau) ,
\ee
which applies for scales $k \ll k_c$. The coefficient $f_h \equiv 1 - (|A_-|/|A_+|)^2$, while $\mathcal{I}(\tau)$ is a time-dependent function which is an integral over the photons Green's functions.

The power spectrum of V-mode polarization is then
\be
\label{result1}
P_{\Theta_V}(k) = \frac{1}{I^2}\frac{1}{2\pi^2} k^3 |\delta V_k|^2 = \frac{\mathcal{N}^2}{I^2 2\pi^2} k^3 ,
\ee
corresponding to a spectral index of V-modes $n_V$, defined by $P_V \propto k^{n_V-1}$, given by
\be
n_{V} = 4 .
\ee
Thus we find a deeply blue spectrum of V-mode polarization. The amplitude of the power spectrum depends sensitively on the parameters $g$, $\vartheta$, $f_h$, and the numerical value of the integral $\mathcal{I}(\tau)$. For an estimate of the amplitude, we turn to the other production mechanism:  the coupling $\phi F \tilde{F}$.

\subsection{Direct production of photons via $\phi F \tilde{F} $}

The preheating production of gauge fields via $\phi F \tilde{F}$ was studied by one of the authors in \cite{McDonough:2016xvu}. The mode functions prepared by inflation are amplified, with the production occurring on a characteristic scale. For modes on much larger length scales, the (scalar) energy density fluctuation after the first oscillation is $k$-independent, with an amplitude that is proportional to the effective background energy density $\langle \rho_A \rangle$ deposited in the gauge field,
\begin{equation}
\label{rhoAkSolvedIR}
|\delta \rho_{Ak}|^2 \, \simeq \,  \frac{\langle \rho_A \rangle^2}{ (2 \xi a_{end} H_{end}) ^3} \, .
\end{equation}
The value of $\langle \rho_A \rangle$ is in turn bounded by backreaction considerations, which ultimately gives for the fluctuations, in the strong-coupling regime,
\bea \label{rhoAfluct}
\delta \rho_{Ak} \, &\sim& \, \frac{V(\phi_{end})}{ (2 \xi  a_{end} H_{end})^{3/2}} \frac{\Lambda_{UV}}{M_{Pl}} \,\,\,\, {\rm for} \,\,\,
\Lambda_{UV} \equiv (\alpha/f)^{-1}< M_{Pl} \nonumber \, . 
\eea
As in the fermionic preheating scenario, this region of parameter space causes preheating to terminate after one production event, such that a maximum polarization asymmetry is achieved. 

In this case the spectrum of super-Hubble V-mode polarization is identical to the spectrum of energy density fluctuations,
\be
\delta V_k = \delta \rho_{Ak} .
\ee
 The power spectrum of V-mode anisotropies at the end of reheating \footnote{We assume that reheating occurs instantaneously after preheating, and that the photons produced during reheating are unpolarized. The total $I$ at the culmination of reheating is proportional to the total energy density of the universe $\rho = 3 M_{Pl}^2 H^2$, and this growth in the intensity $I$ is not accompanied by a growth in $\delta V_k$, as the tachyonic instability is no longer present during perturbative reheating.} is given by
\be
\label{result2}
P_{\Theta_V}(k) =  \frac{1}{2\pi^2} \left( \frac{\Lambda_{UV}}{M_{Pl}}\right)^2 \left( \frac{k}{2 \xi a_{end}H_{end}}\right)^3.
\ee
This is again blue-tilted with a spectral index $n_V = 4$, where $n_V=1$ corresponds to a scale-invariant spectrum. 

The above expression can be written in the parametrized form \eqref{PThetaVForm} as,
\be
P_{\Theta_V}(k) =  {\cal A}_{V} \left( \frac{k}{k_0}\right)^{n_V - 1} \;\;,\;\; n_V=4,
\ee
where ${\cal A}_V \equiv {\cal A}_V(k_0)$ is the amplitude at a reference scale $k_0$ \footnote{Note that the choice of $k_0$ is arbitrary and does not change the physical amplitude of a given $k$-mode. This is analogous to the pivot scale used in the power spectra for scalar and tensor fluctuations  \cite{Ade:2015lrj} }. For $k_0$ that exits the horizon sometime during inflation, such that $k_0 = a_0 H_0$,  the amplitude ${\cal A}_V$ is given by
\be
{\cal A}_{V} = \frac{1}{16 \pi^2 \xi^3}\left( \frac{\Lambda_{UV}}{M_{Pl}}\right)^2 \left( \frac{a_0 H_0}{a_{end}H_{end}}\right)^3 .
\ee
Due to the severe blue-tilt about $k_{end}= a_{end} H_{end}$, the amplitude is suppressed on large scales by a factor $(a_{0}/a_{end})^3 = e^{-3 N_{0}}$, where $N_{0}$ is the number of e-folds of inflation remaining when the mode $k_{0}$ exits the Hubble radius. For the benchmark values of $\xi = \mathcal{O}(1)$, $\Lambda_{UV}\equiv f/\alpha = 10^{-2} M_{Pl}$, the amplitude is given by
\be
\label{result3}
{\cal A}_V  \approx 10^{-7} e^{-3 N_0} .
\ee
From this we see that the severe blue-tilt guarantees a majority of the integrated power will reside in modes that exit the horizon in the last e-fold of inflation.  Choosing the reference scale at $k_0=k_{end}$, the amplitude is given as ${\cal A}_V \approx 10^{-7}$.

 \section{Discussion}
\label{sec:Discussion}
In this work we have found that axion inflation with the standard $\partial_\mu \phi J^{\mu 5}$ and $\phi F \tilde{F}$ couplings produces circular polarization with a spectral index $n_V=4$.  Currently, there has been no detection of $V$, and only upper limits on $C_l ^{VV}$ exist, e.g. as reported by the SPIDER collaboration \cite{Nagy:2017csq} and MIPOL \cite{Mainini:2013mja}. Given this, our work is in a similar spirit to computations of the tensor spectral index $n_T$, as primordial tensor perturbations are in a similar position of not having been observed at all, let alone their spectral index. However, $n_T$ is a remarkably powerful tool for distinguishing models of the early universe: simple single-field inflation models predict $n_T < 0 $, while String Gas Cosmology predicts $n_T >0$ \cite{Brandenberger:2006xi}  \footnote{Blue-tilted super-horizon tensor modes  can also be realized in certain non-minimal inflation models, see e.g.\cite{bluetensorsinflation}, and also in axion inflation coupled to gauge fields \cite{Namba:2015gja}.}. Here we have found that $n_V=4$ is a generic prediction of axion inflation. It would be interesting to construct inflationary models with different values of $n_V$, and in particular $n_V=1$, corresponding to a scale-invariant spectrum of V-mode polarization. It was shown in \cite{Giovannini:2009ru} that a nearly scale-invariant spectrum of V-modes can be generated by  large-scale magnetic fields; it will be very interesting to connect this with models of inflationary magnetogenesis (as reviewed in e.g. \cite{magnetogenesis}, and analysed in \cite{Adshead:2016iae} for axion inflation with the $\phi F \tilde{F}$ coupling we consider here).

The polarization computed here is present at the end of inflation/reheating. However, we have not touched upon the evolution from the end of reheating to the CMB. The evolution to last scattering is described by the Boltzmann equation \cite{Challinor:2000as},
\be
\label{boltzmann}
\dot{V}_{A_l} + \frac{4}{3} \Theta V_{A_{l}} - \frac{l}{2l +1} D_{\langle a_l} V_{A_{l-1} \rangle} + D^b V_{b A_{l}} = - n_e \sigma_T (V_{A_{l}} - \frac{1}{2} V_{a_1} \delta^1 _l) ,
\ee
where $\sigma_T$ is the Thomson cross section, $n_e$ is the free electron density, $\Theta\equiv \nabla_a u^a $ is the volume expansion, and $A_l$ is a string of indices $a_1 .. a_l$. We refer the reader to \cite{Challinor:2000as} for further details on the notation. The above equation (or rather, scalar multipole moments of the above equation) must be incorporated into a CMB Boltzmann solver, such as CAMB, in order to make precise predictions for the $C_{l} ^{VV}$ observed by CMB experiments.  Our results serve as the initial conditions for this analysis. It will be interesting to see if the existing upper limits on V set by MIPOl \cite{Mainini:2013mja} or SPIDER \cite{Nagy:2017csq} can already place constraints on the mechanism discussed here.

The evolution of circular polarization after horizon re-entry was discussed in \cite{Alexander:2016moy}, where it was found that there is an exponential suppression of $V$ due to QED interactions.  At a temperature scale below the mass of the electron, Thomson scattering washes out any net photon helicity due to the large optical depth at this scale. Even with an initial $V/I \simeq 1$, the standard cosmological treatment of the radiation Boltzmann equation could potentially render primordial circular polarization undetectable in the CMB.  

There are, however, potential mechanisms to subvert this exponential decay. As the universe expands, the efficiency at which Thomson scattering can suppress circular polarization diminishes.  Hence, there is a temperature scale $T_c$ below which circular polarization will receive negligible corrections due to Compton scattering, and a window is provided between $T_c$ and last scattering during which sources of circular polarization may be present and detectable in the CMB.  As an example, \cite{Giovannini:2009ru} shows that magnetic fields present in the plasma at last scattering can source circular polarization. Alternatively, a more direct late time production can be found if the axion's velocity, $\dot{\phi}$, is nonzero at late times. Finally, work on cosmic birefringence \cite{Caldwell:2011pu} has suggested that a rotation angle between E and B-mode polarization can arise from a Chern-Simons term.  This rotation angle relies on late time dynamics of the pseudo-scalar field, hence one should also expect a late time production of circular polarization.

Each of these three mechanisms for preventing the decay of $V$ can be described in terms of physical phenomena in the plasma at last scattering.  For a constant magnetic field, the dielectric constant becomes dependent on the helicity of the propagating photon.  The Chern-Simons term causes a relative change in the dispersion relation of the photons.  Finally, the chemical potential in fermions will induce different plasma frequencies for each photon helicity.  In each case, if the mechanism is present in the plasma sufficiently early, the full Compton cross section can conserve photon helicity in interactions, preserving some primordial V-mode polarization. We leave a more detailed description of these phenomena for future work.

There are many other directions for future work that we have not touched upon here. Foremost among this is the analysis of cross-correlation of $V$ with other CMB observables. For example, it is known that $\phi F \tilde{F}$ yields a characteristic tensor mode signal \cite{Namba:2015gja}; it will be interesting to study the cross-correlation of $V$ and $B$ in this model. Such a complete analysis will maximize the information that can be extracted from future CMB experiments, and the constraints on axion inflation that can derived.

\section*{Acknowledgements}

The authors thank Yacine Ali-Ha\"{i}moud, Robert Brandenberger, Elisa Ferreira, Evangelos Sfakianakis, David Spergel, Thomas Tram, and Vincent Vennin, for useful comments and discussions. The authors thank an anonymous referee for many insightful comments and useful suggestions. EM thanks Brown University for hospitality while a portion of this work was completed. EM is supported by The Natural Sciences and Engineering Research Council of Canada (NSERC) via a PGS D fellowship.

\newpage

\appendix

\section{Computation of $\delta V_k$}
\label{appVk}
 We want to study Fourier modes of the stokes parameter $V$. However, since $V\sim \dot{A}^2 \sim \psi^4$ is a composite operator, we have to be careful in how we proceed.

Let's first consider a general (real) operator $\mathcal{O}(x,t)$. This can be split into a background and fluctuation piece via the definition
\be
\mathcal{O}(x,t) = \langle \mathcal{O} \rangle(t) + \delta \mathcal{O}(x,t) ,
\ee
where $<..>$ denotes a classical ensemble average or quantum vacuum expectation value, on super- and sub-Hubble scales respectively. The fluctuation piece can be expanded into plane waves as 
\be
\delta \mathcal{O}(x,t) = \int \frac{{\rm d}^3 k}{(2 \pi)^3} \delta \mathcal{O}_k \, \alpha_k \, e^{i k x} ,
\ee
where $\alpha_k$ are classical random variables, or quantum mechanical annihilation/creation operators, with $\alpha_k= \alpha^* _{-k}$ (which allowed us in the above to combine the positive and negative frequency modes into one term). The mode functions  $\delta \mathcal{O}_k$ are then given
\be
\label{Omodefunctions}
|\delta \mathcal{O}_k |^2 = \int {\rm d}^3 x e^{- i k x} \, \left[ \, \langle  \mathcal{O}(x) \mathcal{O}(0) \rangle - \langle \mathcal{O}\rangle^2 \right] .
\ee
As an illustrative example, one can consider $\mathcal{O}=\phi^2$ for a scalar field $\phi$. In this case, \eqref{Omodefunctions} leads to
\be
|\delta  (\phi \phi)_k |^2 = 2 \int \frac{{\rm d}^3 k'}{(2 \pi)^3} |\phi_{k'}|^2 |\phi_{k-k'}|^2 . 
\ee

For the case of interest for the current work, the mode functions are given by
\be
|\delta V_k|^2 = 2 \int {\rm d}^3 x \, e^{- i k x} \, \left[ \, \langle  V(x) V(0) \rangle - \langle V \rangle^2 \right] 
\ee
Using the expressions of the previous sections, a two-point function $\langle  V(x) V(y) \rangle$ is given in terms of a 4-point function of fermion currents \footnote{Note: this expression already implicitly assumes that only one photon polarization is amplified}, 
\begin{eqnarray}
\langle V(x) V(y) \rangle =  \frac{f_{h}^2}{a^8} \int && \mathrm{d}\eta_1  \mathrm{d}\eta_2 \mathrm{d}\eta_3 \mathrm{d}\eta_4 \; a(\eta_1) a(\eta_2) a(\eta_3) a(\eta_4) \; G'(\eta_1,\tau) {G^*}'(\eta_2,\tau) G'(\eta_3,\tau) {G^*}'(\eta_4,\tau) \nonumber \\
&&\;\;\; \cdot \langle J^{\mu}  (x,\eta_1) J_{\mu} ^\dagger(x,\eta_2) J^{\nu} (y,\eta_3) J_{\nu} ^\dagger (y,\eta_4)  \rangle . 
\end{eqnarray}
where $f_h \equiv 1 - (|A_-|/|A_+|)^2$ is the efficiency of helicity transfer from the $(\pm)$ fermions to the $(\pm)$ circularly polarized photons: if $f_h=1$, then helicity is conserved at every interaction, and only + photons are produced. This, in turn, is an 8-point function of fermions (note that $(\overline{\psi} \gamma_\mu \psi)^\dagger = \overline{\psi} \gamma_\mu \psi$):
\begin{equation}
\langle J^{\mu} (x,\eta_1) J^\dagger _{\mu} (x,\eta_2) J^{\nu} (y,\eta_3) J^\dagger _{\nu} (y,\eta_4)  \rangle = g^4 \langle (\overline{\psi} \gamma^\mu \psi)_{x,\eta_1} (\overline{\psi} \gamma_\mu \psi)_{x,\eta_2} (\overline{\psi} \gamma^\nu \psi)_{y,\eta_3} (\overline{\psi} \gamma_\nu \psi) _{y,\eta_4} \rangle
\end{equation}
This 8-point function can be computed using the fermionic version of Wick's theorem, keeping track of factors of $(-1)$ from shuffling the fermions. We can begin by decomposing it into 4-point functions:
\begin{eqnarray}
\frac{1}{g^4} \langle J^{\mu} (x,\eta_1) J^\dagger _{\mu} (x,\eta_2) J^{\nu} (y,\eta_3) J^\dagger _{\nu} (y,\eta_4)  \rangle =  && \langle (\overline{\psi} \gamma^\mu \psi)_{x,\eta_1} (\overline{\psi} \gamma_\mu \psi)_{x,\eta_2} \rangle \cdot \langle (\overline{\psi} \gamma^\nu \psi)_{y,\eta_3} (\overline{\psi} \gamma_\nu \psi) _{y,\eta_4} \rangle \nonumber \\
&& \;\;\;\; + \langle (\overline{\psi} \gamma^\mu \psi)_{x,\eta_1} (\overline{\psi} \gamma^\nu \psi)_{y,\eta_3} \rangle \cdot \langle (\overline{\psi} \gamma_\mu \psi)_{x,\eta_2} (\overline{\psi} \gamma_\nu \psi) _{y,\eta_4} \rangle \nonumber \\
&& \;\;\;\; + \langle (\overline{\psi} \gamma^\mu \psi)_{x,\eta_1} (\overline{\psi} \gamma_\nu \psi)_{y,\eta_4} \rangle \cdot \langle (\overline{\psi} \gamma_\mu \psi)_{x,\eta_2} (\overline{\psi} \gamma^\nu \psi) _{y,\eta_3} \rangle
\end{eqnarray}
The first term is precisely $\langle V \rangle^2$, leaving only the last two terms to determine $\delta V_k$. Additionally, since the integral is invariant under the exchange $\eta_3 \leftrightarrow \eta_4$, the last two terms will give identical contributions.  Returning to our expression for $\delta V_k$, we now have
\begin{eqnarray} \label{dvk4pt}
|\delta V_k|^2 = 4 g^4 f_h^2 \int {\rm d}^3 x e^{- i k x} \; \frac{1}{a^8} \int && \mathrm{d}\eta_1  \mathrm{d}\eta_2 \mathrm{d}\eta_3 \mathrm{d}\eta_4 \; a(\eta_1) a(\eta_2) a(\eta_3) a(\eta_4) \; G'(\eta_1,\tau) {G^*}'(\eta_2,\tau) G'(\eta_3,\tau) {G^*}'(\eta_4,\tau) \nonumber \\
&&\;\;\; \cdot  \langle (\overline{\psi} \gamma^\mu \psi)_{x,\eta_1} (\overline{\psi} \gamma^\nu \psi)_{y,\eta_3} \rangle \cdot \langle (\overline{\psi} \gamma_\mu \psi)_{x,\eta_2} (\overline{\psi} \gamma_\nu \psi) _{y,\eta_4} \rangle .
\end{eqnarray}
The remaining four-point function can be split into two-point functions using Wick's theorem.  However it is convenient to decompose the four-component fermion $\psi$ into two-component spinors $\varphi, \eta$ since
\begin{equation}
\frac{1}{g}\langle J^{\mu} (x,\eta_1)\rangle = \frac{1}{g}\overline{\sigma}^{\mu ab}\langle J_{ab}\rangle = \overline{\sigma}^{\mu ab} \left(\langle\varphi^\dagger_a \varphi_b\rangle  - \langle\eta^\dagger_a \eta_b\rangle\right) = 0
\end{equation}
becomes automatically imposed.  The product of four-point functions appearing on the second line of \eqref{dvk4pt} can be written as
\begin{equation}
\langle (\overline{\psi} \gamma^\mu \psi)_{x,\eta_1} (\overline{\psi} \gamma_\nu \psi)_{y,\eta_3} \rangle \cdot \langle (\overline{\psi} \gamma_\mu \psi)_{x,\eta_2} (\overline{\psi} \gamma^\nu \psi) _{y,\eta_4} \rangle =\frac{1}{g^4} \langle J^{\mu}(x,\eta_1)J^{\nu}(y,\eta_3)\rangle\cdot\langle J_{\mu}(x,\eta_2)J_{\nu}(y,\eta_4)\rangle
\end{equation}
and the remaining four-point function has the form
\begin{align}
\frac{1}{g^2}\overline{\sigma}^{\mu ab}\overline{\sigma}^{\nu cd}\langle J_{ab}(x,\eta_1)J_{cd}(y,\eta_3)\rangle &=\overline{\sigma}^{\mu ab}\overline{\sigma}^{\nu cd}\langle\left(\varphi^\dagger_a \varphi_b  - \eta^\dagger_a \eta_b\right)_{x,\eta_1}\left(\varphi^\dagger_c \varphi_d  - \eta^\dagger_c \eta_d\right)_{y,\eta_3}\rangle \\
&=\overline{\sigma}^{\mu ab}\overline{\sigma}^{\nu cd}\left(\langle\varphi^\dagger_a \varphi_b\varphi^\dagger_c \varphi_d\rangle-\langle\eta^\dagger_a \eta_b\varphi^\dagger_c \varphi_d\rangle-\langle\varphi^\dagger_a \varphi_b\eta^\dagger_c \eta_d\rangle+\langle\eta^\dagger_a \eta_b\eta^\dagger_c \eta_d\rangle\right).
\end{align}
Again, Wick's theorem can be used to split the fermion four-point function into two-point functions as
\begin{gather}
\langle\varphi^\dagger_a \varphi_b\varphi^\dagger_c \varphi_d\rangle = \langle\varphi^\dagger_a \varphi_b\rangle\cdot\langle\varphi^\dagger_c \varphi_d\rangle- \langle\varphi^\dagger_a \varphi_d\rangle\cdot\langle\varphi_b \varphi^\dagger_c\rangle\\
\langle\eta^\dagger_a \eta_b\varphi^\dagger_c \varphi_d\rangle = \langle\eta^\dagger_a \eta_b\rangle\cdot\langle\varphi^\dagger_c \varphi_d\rangle + \langle\eta^\dagger_a\varphi^\dagger_c\rangle\cdot\langle\eta_b\varphi_d\rangle\\
\langle\varphi^\dagger_a \varphi_b\eta^\dagger_c \eta_d\rangle = \langle\varphi^\dagger_a \varphi_b\rangle\cdot\langle\eta^\dagger_c \eta_d\rangle + \langle\varphi^\dagger_a\eta^\dagger_c\rangle\cdot\langle\varphi_b\eta_d\rangle\\
\langle\eta^\dagger_a \eta_b\eta^\dagger_c \eta_d\rangle = \langle\eta^\dagger_a \eta_b\rangle\cdot\langle\eta^\dagger_c \eta_d\rangle - \langle\eta^\dagger_a \eta_d\rangle\cdot\langle\eta_b \eta^\dagger_c\rangle.
\end{gather}
The first term on the right hand side of these four-point functions will factor to $\langle J_{ab}\rangle\langle J_{cd}\rangle$. The fluctuation piece of the four-point function can then be written as
\begin{align}
\frac{1}{g^2}\langle J^{\mu}(x,\eta_1)J^{\nu}(y,\eta_3)\rangle = 2\overline{\sigma}^{\mu ab}\overline{\sigma}^{\nu cd}&\Big(\langle\varphi^\dagger_a(x,\eta_1) \varphi_d(y,\eta_3)\rangle\cdot\langle\varphi^\dagger_c(y,\eta_3)\varphi_b (x,\eta_1)\rangle \nonumber\\
 &+ \langle\eta^\dagger_{(a}(x,\eta_1)\varphi^\dagger_{c)}(y,\eta_3)\rangle\cdot\langle\varphi_{(d}(y,\eta_3)\eta_{b)}(x,\eta_1)\rangle\rangle\Big) ,
\end{align}
where round brackets $( .. )$ around indices denotes symmetrized indices. The two-point functions appearing above can be written explicitly in terms of fermion mode functions, which have the general form\footnote{Using the relation $\xi^\lambda(\hat{k}) = \xi^{-\lambda}(- \hat{k})$, which follows from the explicit form of the eigenspinors, given in \cite{Koksma:2009tc}.}
\begin{align}
\langle\varphi^\dagger_a(x,\eta_1) \varphi_d(y,\eta_3)\rangle &= \sum_{\lambda}\int\frac{{\rm d}^3 k}{(2 \pi)^3}\left[X^{-\lambda *}_{k}(\eta_1) X^{-\lambda}_{k}(\eta_3)- Y^{\lambda *}_{ k}(\eta_1)Y^{\lambda}_{k}(\eta_3)\right]\xi^\lambda_a({\bf k})\xi^{\lambda\dagger}_d({\bf k})e^{i\kk\cdot(\xx-\yy)},\\
\langle\eta^\dagger_a(x,\eta_1) \varphi^\dagger_c(y,\eta_3)\rangle &= \sum_{\lambda}\int\frac{{\rm d}^3 k}{(2 \pi)^3}\left[Y^{\lambda *}_{k}(\eta_1) X^{\lambda *}_{k}(\eta_3)- Y^{-\lambda *}_{ k}(\eta_3)X^{-\lambda *}_{k}(\eta_1)\right]\xi^\lambda_a({\bf k})\xi^{\lambda\dagger}_c({\bf k})e^{i\kk\cdot(\xx-\yy)}.
\end{align}
On large scales we can expand the fermion mode functions as
\begin{eqnarray}
X_k^+(k\tau) &&= -\left(\frac{m_{\psi}}{H}\Gamma(-2i\vartheta)\right)(1-i)\;2^{-1+i\vartheta}\; e^{ik\tau+\pi\vartheta}(-k\tau)^{i\vartheta},\\
X_k^-(k\tau) &&= -(1+i)\;2^{-1+i\vartheta}\; e^{-ik\tau}(-k\tau)^{i\vartheta},\\
Y_k^{+*}(k\tau) &&= -(1+i)\;2^{-1-i\vartheta}\; e^{-ik\tau}(-k\tau)^{-i\vartheta},\\
Y_k^{-*}(k\tau) &&= -\left(\frac{m_\psi}{H}\Gamma(2i\vartheta)\right)(1-i)\;2^{-1-i\vartheta}\; e^{ik\tau-\pi\vartheta}(-k\tau)^{-i\vartheta}.
\end{eqnarray}
We then define the quantities:
\begin{gather}
\mathcal{A}_k^{\lambda}(\eta_i,\eta_j) = \left[X^{-\lambda *}_{k}(\eta_i) X^{-\lambda}_{k}(\eta_j)- Y^{\lambda *}_{ k}(\eta_i)Y^{\lambda}_{k}(\eta_j)\right],\\
\mathcal{B}_k^{\lambda}(\eta_i,\eta_j) = \left[Y^{\lambda *}_{k}(\eta_i) X^{\lambda *}_{k}(\eta_j)- Y^{-\lambda *}_{ k}(\eta_j)X^{-\lambda *}_{k}(\eta_i)\right]. 
\end{gather}
In general, neither $A_k^\lambda, B_k^\lambda$ are nonzero, however we can order them for small $m_\psi/H$,
\begin{equation}
\mathcal{A}^+ \sim \mathcal{O}(1) \;\;\;,\;\;\; \mathcal{B}^\pm \sim \mathcal{O}\left(\frac{m}{H}\right)\;\;\;,\;\;\; \mathcal{A}^- \sim \mathcal{O}\left(\frac{m^2}{H^2}\right)
\end{equation}
To lowest order in $m_\psi/H$,  the fermion four-point function takes the form
\begin{equation}
\frac{1}{g^2}\langle J^{\mu}(x,\eta_1)J^{\nu}(y,\eta_3)\rangle = 2\overline{\sigma}^{\mu ab}\overline{\sigma}^{\nu cd}\int\frac{{\rm d}^3 k_1}{(2 \pi)^3}\frac{{\rm d}^3 k_2}{(2 \pi)^3}\mathcal{A}_{k_1}^+(\eta_1,\eta_3)\mathcal{A}_{k_2}^+(\eta_3,\eta_1)\xi^+_a({\bf k}_1)\xi^{+\dagger}_d({\bf k}_1)\xi^+_c({\bf k}_2)\xi^{+\dagger}_b({\bf k}_2)e^{i(\kk_1-\kk_2)\cdot(\xx-\yy)}
\end{equation}
where the next order correction is $\mathcal{O}(m/H)^2$.  Therefore, the product appearing on the second line of \eqref{dvk4pt} can be written in the form (now dropping the $+$ superscript from the $\xi$'s):
\begin{align}
\frac{1}{g^4} \langle J^{\mu}(x,\eta_1)J^{\nu}(y,\eta_3)\rangle\cdot\langle J_{\mu}(x,\eta_2)J_{\nu}(y,\eta_4)\rangle = 4 \overline{\sigma}^{\mu ab}\overline{\sigma}^{\nu cd}\overline{\sigma}_\mu^{rs}&\overline{\sigma}_\nu^{tu}\int_0 ^{\vartheta a H}   \frac{{\rm d}^3 k_1}{(2 \pi)^3} \, \frac{{\rm d}^3 k_2}{(2 \pi)^3} \, \frac{{\rm d}^3 k_3}{(2 \pi)^3}  \frac{{\rm d}^3 k_4}{(2 \pi)^3}\\
&\times\Bigg[\exp[i(\kk_1-\kk_2+\kk_3-\kk_4)\cdot(\xx-\yy)]\nonumber\\
&\times \mathcal{A}_{k_1}^+(\eta_1,\eta_3)\mathcal{A}_{k_2}^+(\eta_3,\eta_1)\mathcal{A}_{k_3}^+(\eta_2,\eta_4)\mathcal{A}_{k_4}^+(\eta_4,\eta_2)\nonumber\\
&\times\xi_a({\bf k}_1)\xi^{\dagger}_d({\bf k}_1)\xi_c({\bf k}_2)\xi^{\dagger}_b({\bf k}_2)\xi_r({\bf k}_3)\xi^{\dagger}_u({\bf k}_3)\xi_t({\bf k}_4)\xi^{\dagger}_s({\bf k}_4)\Bigg]\nonumber
\end{align}
where the function $\mathcal{A}_{k}^+(\eta_1,\eta_2)$ can be explicitly written as
\begin{equation}
\mathcal{A}_{k}^+(\eta_1,\eta_2) = i\sin\left(k(\eta_1-\eta_2)\right)\exp\left[-i\vartheta\log\left(\frac{\eta_1}{\eta_2}\right)\right] .
\end{equation}

Finally, the mode functions of the circular polarization (using the inflationary fermion mode functions and taking the lowest order in mass) are given by:
\begin{align}
|\delta V_k|^2 = &16 g^4 f_h^2  \frac{1}{a^8(\tau)}\int \mathrm{d}\eta_1  \mathrm{d}\eta_2 \mathrm{d}\eta_3 \mathrm{d}\eta_4 \; a(\eta_1) a(\eta_2) a(\eta_3) a(\eta_4) \; G'(\eta_1,\tau) {G^*}'(\eta_2,\tau) G'(\eta_3,\tau) {G^*}'(\eta_4,\tau)\\
&\times \int {\rm d}^3 x \int_0 ^{\vartheta a H}   \frac{{\rm d}^3 k_1}{(2 \pi)^3} \, \frac{{\rm d}^3 k_2}{(2 \pi)^3} \, \frac{{\rm d}^3 k_3}{(2 \pi)^3}  \frac{{\rm d}^3 k_4}{(2 \pi)^3}\Bigg[\exp[i(\kk_1-\kk_2+\kk_3-\kk_4 -\kk )\cdot\xx]\nonumber\\
&\times\sin\left(k_1(\eta_1-\eta_3)\right)\sin\left(k_2(\eta_1-\eta_3)\right)\sin\left(k_3(\eta_2-\eta_4)\right)\sin\left(k_4(\eta_2-\eta_4)\right)\nonumber\\
&\times\xi(k_1) \overline{\sigma}^\mu \xi^\dagger(k_2) \cdot \xi(k_2) \overline{\sigma}^\nu \xi^\dagger(k_1) \cdot \xi(k_3) \overline{\sigma}_\mu \xi^\dagger(k_4) \cdot \xi(k_4) \overline{\sigma}_\nu \xi^\dagger(k_3)\Bigg]\nonumber .
\end{align}
We can now perform the $x$-integration and one of the $k_i$-integrations. If we choose $i=4$, this sets $\kk_4= \kk_1 -\kk_2 +\kk_3 - \kk$. For $k/aH \ll 1 $, the remaining $k$-integrals are dominated by the upper bound $k_i = \vartheta a H$.
\begin{align}
|\delta V_k|^2 \approx &\frac{16 g^4 f_h^2}{a^8(\tau)}  \left( \vartheta a H\right)^9  \int \mathrm{d}\eta_1  \mathrm{d}\eta_2 \mathrm{d}\eta_3 \mathrm{d}\eta_4 \; a(\eta_1) a(\eta_2) a(\eta_3) a(\eta_4) \; G'(\eta_1,\tau) {G^*}'(\eta_2,\tau) G'(\eta_3,\tau) {G^*}'(\eta_4,\tau)\\
&\times \sin\left(\vartheta a H(\eta_1-\eta_3)\right)\sin\left(\vartheta a H(\eta_1-\eta_3)\right)\sin\left(\vartheta a H(\eta_2-\eta_4)\right)\sin\left(\vartheta a H(\eta_2-\eta_4)\right)\nonumber\\
&\times  \int_{|k_i|= \vartheta a H} \mathrm{d}\theta_1 \mathrm{d}\phi_1 \mathrm{d}\theta_2 \mathrm{d}\phi_2 \mathrm{d}\theta_3 \mathrm{d}\phi_3 \xi(\theta_1,\phi_1) \overline{\sigma}^\mu \xi^\dagger(\theta_2, \phi_2) \cdot \xi(\theta_2,\phi_2) \overline{\sigma}^\nu \xi^\dagger(\theta_1, \phi_1) \cdot \xi(\theta_3, \phi_3) \overline{\sigma}_\mu \xi^\dagger(\kk_4) \cdot \xi(\kk_4) \overline{\sigma}_\nu \xi^\dagger(\theta_3, \phi_3)\nonumber ,
\end{align}
where the final line is an integral over the angular variables of the $k_i$ at $|k_i|=\vartheta a H$, (recall that $\xi_\lambda(\kk)$ depends only on $\hat{k}$), and $\kk_4 \approx \kk_1 - \kk_2 + \kk_3$ is evaluated at $|k_1|=|k_2|=|k_3|=\vartheta a H$ . Again we note that all $\xi$'s appearing above are $\xi^+$. This result has the schematic form,
\be
|\delta V_k|^2 =  \frac{16 g^4 f_h^2}{a^8(\tau)}  \left( \vartheta a H\right)^9  \mathcal{I}(\tau)
\ee
where $\mathcal{I}(\tau)$ is integral over Green's functions given above, and the angular integral over the helicity eigenspinors.
As per the discussion in Section IV, the above $|\delta V_k|^2$ (valid on large scales) is $k$-independent.

\section{Competition of couplings}

We would like to argue that there exists a regime in which the inflaton-fermion interaction is dominant for the production of circular polarization, while inflaton-gauge preheating has a subleading role.

Consider the gauge field with the usual QED interaction in addition to a Pontryagin coupling with the inflaton.  There is also a derivative coupling between the chiral fermion current and the axion.  The action for the gauge field is given by
\begin{equation}
S = \int {\rm d}^4 x \sqrt{- g} \left[-\frac{1}{4}F_{\mu\nu}F^{\mu\nu} + A_{\mu} J^{\mu} + \frac{\alpha}{f}\phi F_{\mu\nu}\tilde{F}^{\mu\nu} \right] .
\end{equation}
Then, the equation of motion for the gauge field with different helicities can be written as
\begin{equation}
\left(\partial_{\tau}^2 + k^2 \pm \frac{\alpha}{f}\frac{\dot{\phi}}{a(\tau)}k\right)A^{\pm}_k(\tau) = -\frac{1}{a(\tau)}J^{\pm}_k(\tau)
\end{equation}
where the fermions have been rescaled, as in \cite{Adshead:2015kza}.  We are interested in the magnitude of the contributions of the two interaction term with the gauge field. By defining $\xi =\frac{\alpha}{f}\frac{\dot{\phi}}{H}$, then we can rewrite the equation of motion as
\begin{equation}
\left(\partial_{\tau}^2 + k^2\right)A^{\pm}_k(\tau) = -\frac{1}{a(\tau)}\left(J^{\pm}_k(\tau)\pm \xi HkA^{\pm}_k(\tau)\right)
\end{equation}
and we are interested in calculating the relative magnitudes of the two terms on the right side.

If we impose that the QED interaction between the gauge field and fermions dominates during inflation, we have the general solution to the gauge field equation of motion as
\begin{equation}
A^{\pm}_k(\tau) \sim i\int\frac{\dd{\eta}}{a(\eta)}G_k(\eta,\tau)J^{\pm}_k(\eta)
\end{equation}
where $G_k$ is the Green's function and we omit the background solution as it is assumed small.  The Green's function has been found \cite{Alexander:2016moy} to be
\begin{equation}
G_k(\eta,\tau) = \frac{-i}{k}\sin \left(k(\tau-\eta)\right)\theta(\tau-\eta)
\end{equation}
Then, the condition we must satisfy becomes
\begin{equation}
|J^{\pm}_k(\tau)|^2 \geq \xi^2 H^4\left|\int\dd{\eta}\eta\sin \left(k(\tau-\eta) \right)J^{\pm}_k(\eta)\right|^2
\end{equation}
To find an upper bound on the integral, we assume that the growth of the current is slower than the change in comoving Hubble radius $aH$.  Then, the integrand has an envelope that is monotonically decreasing, so the integral is dominated when the current is turned on at the initial time, some $\tau_i$.  Hence,
\begin{equation}
|J^{\pm}_k(\tau)|^2 \geq \xi^2 H^4 \tau_i^2 \tau^2 |J^{\pm}_k(\tau_i)|^2
\end{equation}
 We then have the condition on the coupling strength for the $\phi F\tilde{F}$ term as
\begin{equation}
|\xi| \leq a(\tau_i)a(\tau)\frac{|J^{\pm}_k(\tau)|}{|J^{\pm}_k(\tau_i)|} \ll 1. \label{Constr1}
\end{equation}
The direct coupling between the axion and the gauge field must remain small.  The initial time cannot be small (near the end of inflation) as this would contradict the statement that the background gauge field is small, since the field would evolve under the Pontryagin term for a long period of time during inflation.  Intuitively, this is a statement that the direct decay rate of the axion into the photons must be small so that the preferred decay mode is through fermions.

Then, if the Pontryagin term is to dominate after the end of inflation, we need to satisfy the equation
\begin{equation}
|J^{\pm}_k(\tau)|^2 \leq (\xi Hk)^2|A^{\pm}_k(\tau)|^2.
\end{equation}
Here, we know the field will grow due to tachyonic instabilities, and have generally exponential growth from some initial value.
\begin{equation}
|A^{\pm}_k(\tau)|^2 = \exp(\lambda\tau)|A^{\pm}_k(\tau_0)|^2
\end{equation}
where $\tau_0$ is given at the end of inflation.  We have already that the gauge field at the end of inflation should behave as
\begin{equation}
|A^{\pm}_k(\tau_0)|^2 = \left(\frac{H}{k}\right)^2\tau_i^2 \tau_0^2 |J^{\pm}_k(\tau_i)|^2.
\end{equation}
The condition that needs to be satisfied becomes:
\begin{equation}
|J^{\pm}_k(\tau)|^2 \leq \left(\frac{\xi}{a(\tau_i)a(\tau_0)}\right)^2 |J^{\pm}_k(\tau_i)|^2 e^{\lambda \tau}.
\end{equation}
Since the previous condition on $\vartheta$, given by equation (\ref{Constr1}), should be saturated around the end of inflation, the new condition becomes
\begin{equation}
|J^{\pm}_k(\tau)|^2 \leq |J^{\pm}_k(\tau_0)|^2 e^{\lambda\tau}
\end{equation}
Hence, the exponential growth factor has a bound given by
\begin{equation}
\lambda \geq \frac{2}{\tau}\ln\left(\frac{|J^{\pm}_k(\tau)|}{|J^{\pm}_k(\tau_0)|}\right) \label{Consrt2}
\end{equation}
Note, although $\tau$ is defined to be the time since the end of inflation, there should be some finite time for the phase transition near the end of inflation.  Therefore, these considerations must take place some finite $\tau$ after the end of inflation.

The weak coupling of $\xi$ may end up complicating this calculation.  For weak coupling, there should be an extended period of reheating where the axion will undergo many oscillation in its potential.  This will cause the ratio of circular polarization to total intensity of light to diminish after each successive oscillation.  Furthermore, the exponential factor $\lambda$ will generally depend on the $k$ value that is being amplified.  As a result, there will be some cutoff in $k$ where this condition will no longer be satisfied.  In general, this will favor the large $k$ values, where we do not expect a large generation of circular polarization.  Taking the solution for $\lambda$ from \cite{McDonough:2016xvu},
\begin{equation}
\lambda_k = (3.6\times 10^{-3})\left(\frac{k}{\Lambda}\right)^\frac{1}{2}m_{pl}
\end{equation}
where $\frac{1}{\Lambda}=\frac{\xi}{4\sqrt{6}m_{pl}}$.  From this, equation (\ref{Consrt2}) becomes a constraint on the amount of time needed for the phase transition, given by the minimum allowed value of
\begin{equation}
\tau_{min} \geq \frac{2}{\lambda_k}\ln\left(\frac{|J^{\pm}_k(\tau)|}{|J^{\pm}_k(\tau_0)|}\right) \sim \left(10^{9} \text{ s}\right)\left(\frac{m_{pl}}{\xi k}\right)^{\frac{1}{2}}
\end{equation}
The minimum desired value of $k$ then sets the transition time. The smaller value the $k$, the longer the transition will take and the assumption that the Pontryagin term will dominate during reheating no longer becomes valid.  Therefore, it is not valid to produce circular polarization during inflation through a fermion chiral current and amplify the gauge field during reheating though the Pontryagin term.  Based on these arguments, we will consider separately the case of preheating where the fermion-inflaton interaction dominates and when the Pontryagin-inflaton interaction  dominates.

\end{document}